# Optimization of fipronil removal via electro-Fenton using a carbon cloth air-diffusion electrode

Caio Machado Fernandes[1,2*], Gabriel A. Cerrón-Calle[2], Enric Brillas[3],

Mauro C. Santos[1], Sergi Garcia-Segura[2*]

*[1]Laboratório de Eletroquímica e Materiais Nanoestruturados, Centro de Ciências Naturais e Humanas, Universidade Federal do ABC, Santo André, SP, 09210-170, Brasil*

*[2]Nanosystems Engineering Research Center for Nanotechnology-Enabled Water Treatment, School of Sustainable Engineering and the Built Environment, Arizona State University, Tempe, AZ, 85287, USA*

*[3]Laboratori d'Electroquímica dels Materials i del Medi Ambient, Departament de Ciència de Materials i Química Física, Facultat de Química, Universitat de Barcelona, Martí i Franquès 1-11, 08028 Barcelona, Spain*

\*Corresponding author e-mail: sgarcias@asu.edu (S. Garcia-Segura)

\*Corresponding author e-mail: cmacha13@asu.edu (C. Machado Fernandes)

**Abstract**

The electro-Fenton (EF) process using a boron-doped diamond (BDD) anode and a carbon cloth air-diffusion cathode was optimized for efficient fipronil degradation. The system achieved high $H_2O_2$ electrogeneration with close to 80% current efficiency operating between 10 and 50 mA cm$^{-2}$, with hydroxyl radicals formed from BDD oxidation and Fenton's reaction that drive pollutant decay. Optimum conditions were found with 0.50 mM $Fe^{2+}$ catalyst at pH = 3.0 and 30 mA cm$^{-2}$, yielding almost complete removal for 20 mg L$^{-1}$ fipronil in 60 min, 85% removal for 10 mg L$^{-1}$, and a robust performance down to 1 mg L$^{-1}$, thus reflecting real-world applicability. The system demonstrated excellent reusability and stability over seven consecutive runs. The evolution of final short-chain linear carboxylic acids like acetic, fumaric, formic, oxalic, and oxamic, and inorganic ions like F$^-$, Cl$^-$, and $NH_4^+$, was determined. This study highlights the EF process as a highly efficient, energy-balanced, and stable solution for fipronil degradation in water treatment.

*Keywords:* Air-diffusion electrode; Boron-doped diamond anode; Electrochemical advanced oxidation processes; Pesticides; Persistent Organic Pollutants (POP); Water treatment.

# 1. Introduction

Ensuring access to clean water is a fundamental goal of sustainable development [1]. Although water quality is crucial for the health and survival of all living beings, in recent decades, human activities have compromised this essential resource by releasing toxic and non-biodegradable organic pollutants into aquatic ecosystems and environments [2, 3].

Chemical pesticides have been used by farmers to protect crops from rising pest populations. The widespread application of pesticides in agriculture has propitiated the significant environmental accumulation of these recalcitrant pollutants. The high solubility of pesticides in water propitiate infiltrations in soils and water bodies, their spreading through agricultural groundwaters, and eventually resulting in trace pesticide concentrations reaching rivers with the consequent risk to the environment and human health [4-6].

Fipronil (($\pm$)-5-amino-1-(2,6-dichloro-$\alpha,\alpha,\alpha$-trifluoro-p-tolyl)-4-trifluoromethylsulfinylpyrazole-3-carbonitrile, $C_{12}H_4Cl_2F_6N_4OS$, see chemical structure in Fig. 1, $M = 437.14$ g mol$^{-1}$) is a phenylpyrazole insecticide and pesticide used in various agricultural and horticultural practices. The strong affinity of fipronil for soils and sediments leads to a major environmental challenge with widespread contamination. Once in water bodies, fipronil has severe environmental and health impacts. Exposure of humans to fipronil can cause headaches, seizures, paresthesia, pneumonia, and eventually death. Moreover, fipronil contamination can potentially lead to both environmental damage and disruptions in physiological functions across various species [7-12].

The most well-known incident of fipronil contamination has been the "poisonous egg" scandal, which was first detected in the Netherlands but was identified to be quickly spreading worldwide [13]. Despite the ban on fipronil use in China since 2009, residues continue to be found in fruits and vegetables [14]. Fipronil contamination has been detected in groundwater in Switzerland [15], the Elbe River in Germany [16], and water sources in Vietnam [17]. Several researchers highlight that the presence of fipronil is still a big concern in the USA [18-21]. In Brazil, alarming concentrations of this pesticide have been reported in the Aquidauana River, significantly exceeding maximum contaminant levels and compromising the use of safe water [22]. Furthermore, studies in São Paulo revealed fipronil in 62% of freshwater samples, with concentrations surpassing both national and international safety standards for aquatic life [23].

Conventional water treatment methods do not effectively remove pesticides and often fail to comply with stringent regulations, underscoring the need for alternative solutions. Electrified processes may offer a promising decentralized approach to eliminating persistent pesticides. Powerful electrochemical advanced oxidative processes (EAOPs) are based on the *in situ* electrogeneration of highly reactive oxygen species (ROS) like hydroxyl radical ($^\bullet OH$). This radical is the second strongest oxidant known, which is able to fully degrade and mineralize organic pollutants [24-28].

Electrochemical oxidation (ECO) is perhaps the most studied EAOP. During ECO, organics are destroyed by the physisorbed hydroxyl radical (M($^\bullet OH$)) originating at an anode M with high overpotential for oxygen evolution from water oxidation reaction (1) [28,29]. A previous study on the ECO process of 10 mg $L^{-1}$ fipronil solutions employing a boron-doped diamond (BDD) anode and a stainless-steel cathode revealed near 90% abatement after 360 min of electrolysis with a pseudo-first-order rate constant ($k_1$) of 7.2 × $10^{-3}$ $min^{-1}$ at 20 mA $cm^{-2}$ [29].

$$M + H_2O \rightarrow M(^\bullet OH) + H^+ + e^- \qquad (1)$$

The application of electrochemically-driven Fenton processes may accelerate fipronil degradation because they combine the action of heterogeneous M($^\bullet OH$) generated on the anode surface with homogeneous •OH produced in the bulk from Fenton's reaction (2). In electro-Fenton (EF), the required $H_2O_2$ is electrochemically generated in situ at the cathode from $O_2$ reduction via reaction (3) [30-32]. Furthermore, the electrochemical system accelerates the continuous regeneration of the $Fe^{2+}$ catalyst from $Fe^{3+}$ reduction by reaction (4), much faster than the sluggish Fenton-like reaction (5). The faster electro-regeneration of the catalyst decreases the iron concentration needed when compared to conventional chemical Fenton processes while enhancing the sustainability and cost-effectiveness of the process [30-33].

$$H_2O_2 + Fe^{2+} \rightarrow Fe^{3+} + {}^\bullet OH + OH^- \qquad (2)$$

$$O_2 + 2H^+ + 2e^- \rightarrow H_2O_2 \qquad (3)$$

$$Fe^{3+} + e^- \rightarrow Fe^{2+} \qquad (4)$$

$$Fe^{3+} + H_2O_2 \rightarrow Fe^{2+} + {}^\bullet O_2H + H^+ \qquad (5)$$

With respect to the cathode choice, carbon cloth is a highly suitable material for use as a gas diffusion electrode (GDE) due to its excellent electrical conductivity, high surface area, and porous structure, which facilitate efficient oxygen diffusion and reduction. Its three-dimensional architecture enhances the mass transfer of $O_2$, enabling effective *in situ* electrogeneration of $H_2O_2$ via the 2-electron oxygen reduction reaction (ORR). Additionally, carbon cloth is chemically stable under acidic conditions, has the ability to sustain high current efficiencies for $H_2O_2$ production, and is mechanically robust and cost-effective, making it an ideal candidate for scaling up electro-Fenton processes [34,35].

This study aims to optimize the EF degradation of fipronil by leveraging a carbon cloth air-diffusion cathode. Once characterized by this electrode's ability for $H_2O_2$ electrogeneration, the effects of pH, current density (*j*), and pesticide and $Fe^{2+}$ contents on fipronil abatement have been examined. The reusability of the optimum system has been found, and the evolution of final carboxylic acids has been determined. This work contributes to the development of an effective and sustainable EF approach for removing persistent organic pollutants from water.

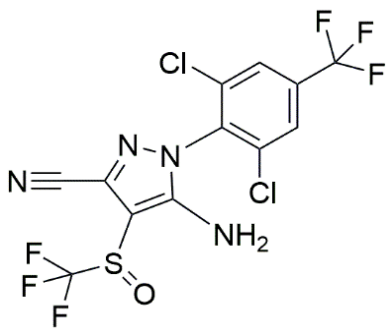

**Fig. 1**. Chemical structure of fipronil.

## 2. Materials and methods

*2.1. Chemicals*

Analytical standard fipronil (PESTANAL®) was supplied by Sigma-Aldrich. A stock solution of this pesticide (1000 mg L$^{-1}$) in analytical methanol was prepared and stored in the refrigerator to be further added to nano-pure water from a Millipore Milli-Q system with a resistivity of 18.2 MΩ cm for the preparation of studied solutions. Anhydrous $Na_2SO_4$ used as supporting electrolyte and $FeSO_4 \cdot 7H_2O$ used as catalyst were of analytical grade purchased from Sigma-Aldrich. Standard carboxylic acids were of analytical grade provided by Sigma-Aldrich. Analytical $H_2SO_4$ supplied by Sigma-Aldrich was added to the solutions for pH adjustment. All the other chemicals were of analytical or high-performance liquid chromatography (HPLC) grade provided by Sigma-Aldrich.

*2.2. Electrochemical system*

The electrochemical experiments were conducted in a conventional 250 mL undivided cell containing 150 mL of solution. The solution treated was kept under continuous stirring at 200 rpm using a magnetic stirrer to ensure uniform mixing and efficient transport of the pesticide towards/from the electrode surfaces. A 10 cm$^2$ BDD electrode served as the anode and a 2.5 cm² carbon cloth air-diffusion electrode as the cathode. The BDD anode was purchased from NeoCoat, whereas the cathode was from the Fuel Cell Store (College Station, TX, USA). The cathode (Woven Carbon Fiber Cloth, 410 microns, Electrical Resistivity (through plane) < 13 mΩcm², PTFE treated) was mounted onto a PVC homemade gas diffusor configuration and continuously electrogenerated $H_2O_2$ from reaction (3) during electrolysis by injecting an airflow rate of 2.0 L min$^{-1}$ through its inner face with an air pump.

All solutions were prepared with 0.020 M $Na_2SO_4$ as a supporting electrolyte. The effect of operating variables was studied considering pH values between 1.0 and 7.0, pesticide concentrations between 1 and 20 mg L$^{-1}$, and catalytic $Fe^{2+}$ dosages between 0.25 and 1.00 mM. The solution pH was adjusted with 1.00 M $H_2SO_4$.

Electrolysis experiments were performed under galvanostatic conditions at a constant *j* with respect to the cathode ranging from 10 to 50 mA cm$^{-2}$ (current from 25 to 125 mA) supplied by a TENMA 72–2720 DC potentiostat/galvanostat as power supply. Aliquots of 1.5 mL were periodically collected during electrolysis at predefined time intervals for subsequent analysis of fipronil degradation. All measurements were conducted in duplicate to ensure the reproducibility and reliability of the results and the error bars obtained with a 95% confidence interval are shown in the figures. Prior to electrolysis, the electrodes were cleaned and activated by polarization in the cell within a 0.020 M $Na_2SO_4$ solution at *j* = 100 mA cm$^{-2}$ for 180 min.

*2.3. Analytical methods*

The $H_2O_2$ electrogenerated by the carbon cloth air-diffusion cathode was quantified following the colorimetric method based on the formation of the yellowish complex Ti(IV)-$H_2O_2$ and measuring its absorbance at $\lambda$ = 408 nm using a DR6000 UV-Vis spectrophotometer (Hach, Ames, IA, USA) [36]. The current efficiency of the electrogeneration of $H_2O_2$ was estimated according to Faraday's law as follows:

$$\% \text{ Current efficiency} = \frac{[H_2O_2]\, n\, F\, V_s}{I\, t\, M} \times 100 \quad (6)$$

where [$H_2O_2$] is the experimental $H_2O_2$ concentration (in mg L$^{-1}$), *n* is the number of electrons involved in reaction (3) (*n* = 2), *F* is the Faraday constant (96487 C.mol$^{-1}$), $V_s$ is the solution volume (in L), *I* is the applied current (in A), *t* is the electrolysis time (in s), and *M* is the molecular mass of $H_2O_2$ (34 g mol$^{-1}$).

The decay of fipronil concentration over time was monitored using a Waters 2695 HPLC system. It was fitted with a Waters Symmetry C18 column (4.6 mm × 75 mm, 3.5 μm) coupled to a Waters 2998 photodiode array detector set at $\lambda$ = 278 nm. A sample volume of 20 μL was injected into the liquid chromatograph and eluted with an 80:20 (v/v) acetonitrile/water mixture that circulated under isocratic conditions at a flow rate of 0.3 mL min$^{-1}$. The peak of fipronil appeared at a retention time of 4.7 min, but no additional peaks related to by-products were recorded [29]. To determine the concentration of final carboxylic acids, the same HPLC system fitted with a Bio-Rad Aminex HPX-87H (7.8 mm × 300 nm) column at 35 ºC and the detector selected at $\lambda$ = 210 nm was used. A sample volume of 20 μL was also injected and eluted with 4.0 mM $H_2SO_4$ at a flow rate of 1.0 mL min$^{-1}$. The retention times were 5.8 min for oxalic acid, 7.2 min for acetic acid, 11.8 min for oxamic acid, 13.5 min for formic acid, and 18.2 min for fumaric acid.

The concentration of $F^-$ and $Cl^-$ ions released during the degradation process was quantified by ionic chromatography using a Shimadzu Prominence Ion Chromatograph with a Shodex IC SI-90G column, coupled with a Shimadzu Detector SPD-40V. The $NH_4^+$ concentration was determined with a colorimetric method using a HACH DR6000 UV Vis spectrophotometer, and the absorbance of samples was taken at $\lambda$ = 655 nm from a solution containing 1 M NaOH, 5 % salicylic acid, 5 % sodium citrate, 0.05 M NaClO, and 1 % sodium nitroferricyanide.

The following pseudo-first-order kinetic equation was checked to analyze the fipronil concentration decay:

$$\ln\left(\frac{c_0}{c}\right) = k_1 t \quad (7)$$

where $c_0$ and $c$ are the initial pesticide concentration and that of time $t$, respectively, and $k_1$ is the corresponding pseudo-first-order rate constant.

The electrical energy per order (EE/O, in kWh m$^{-3}$ order$^{-1}$) was defined as the energy consumption related to the reduction of the fipronil content by one order of magnitude in a given unit volume. It was calculated as follows [37]:

$$\text{EE/O} = \frac{E_{\text{cell}} I t}{V_s \log(c_0/c)} \quad (8)$$

where $E_{\text{cell}}$ is the cell potential (in V), $I$ is the current (in A), and $t$ is the time (in h). If Eq. (7) is verified, it can be expressed as log ($c_0$ /$c$) = 0.4343 $k_1$ $t$, with $k_1$ given in h$^{-1}$, and hence, Eq. (8) can be re-written as:

$$EE/O = \frac{E_{\text{cell}} I t}{0.4343 V_s k_1} \quad (9)$$

## 3. Results and discussion

### 3.1. Assessing effective electrogeneration of hydrogen peroxide

Fig. 2a illustrates the change of accumulated $H_2O_2$ with electrolysis time using a BDD/carbon cloth air-diffusion cell as function of applied $j$ varying from 10 to 50 mA cm$^{-2}$. A progressive increase of $H_2O_2$ concentration can be observed at more extended electrolysis times and higher $j$, reaching 81, 170, 235, 298, and 359 mg L$^{-1}$ in 60 min for 10, 20, 30, 40, and 50 mA cm$^{-2}$, respectively. This trend is consistent with the expected acceleration of reaction (3) when increasing the number of electrons delivered per unit of time (higher $j$), which results in larger yields of $H_2O_2$ as long as there is available $O_2$ to be reduced [38-41].

Fig. 2b highlights that the a similar percentage of current efficiency for $H_2O_2$ accumulation, around 80%, was found for all the $j$ values. This high current efficiency demonstrated that under the present operating conditions, the system is operating under limiting current, i.e., the air-diffusion electrode ensures effective gas delivery giving rise to an environment saturated with $O_2$ at the electrode interface. The capability of sustaining the continuous production of $H_2O_2$ on site is essential to warrant the maximum production of homogeneous •OH from reaction (2) when the EF process is applied [42,43]. The percentage of current efficiency did not reach a 100% value at the beginning of the assays, which can be related to the parallel 4-electron reduction of $O_2$ to $H_2O$ as a competitive reduction reaction. Moreover, $H_2O_2$ may slightly decay over time due to possible oxidation at the anode following reaction (10) or $H_2O_2$ decomposition according to general expression (11). For example, its value dropped from 88.5% at 10 min to 67.8% at 60 min for $j=$ 50 mA cm$^{-2}$. The more significant decay observed at higher applied $j$ supports the hypothesis of anodic oxidation of $H_2O_2$ as the primary parasitic reaction during electrogeneration. It can be related to an increase in the rate of the oxidation of $H_2O_2$ at the BDD anode by reaction (10) due to its gradual accumulation in the medium [44,45]. Note that $H_2O_2$ oxidation yields hydroperoxyl radical ($HO_2^{\bullet}$), a weaker reactive oxygen species ($E^{\circ}$ = 1.44 V/SHE) than $^{\bullet}OH$ ($E^{\circ}$ = 2.8 V/SHE).

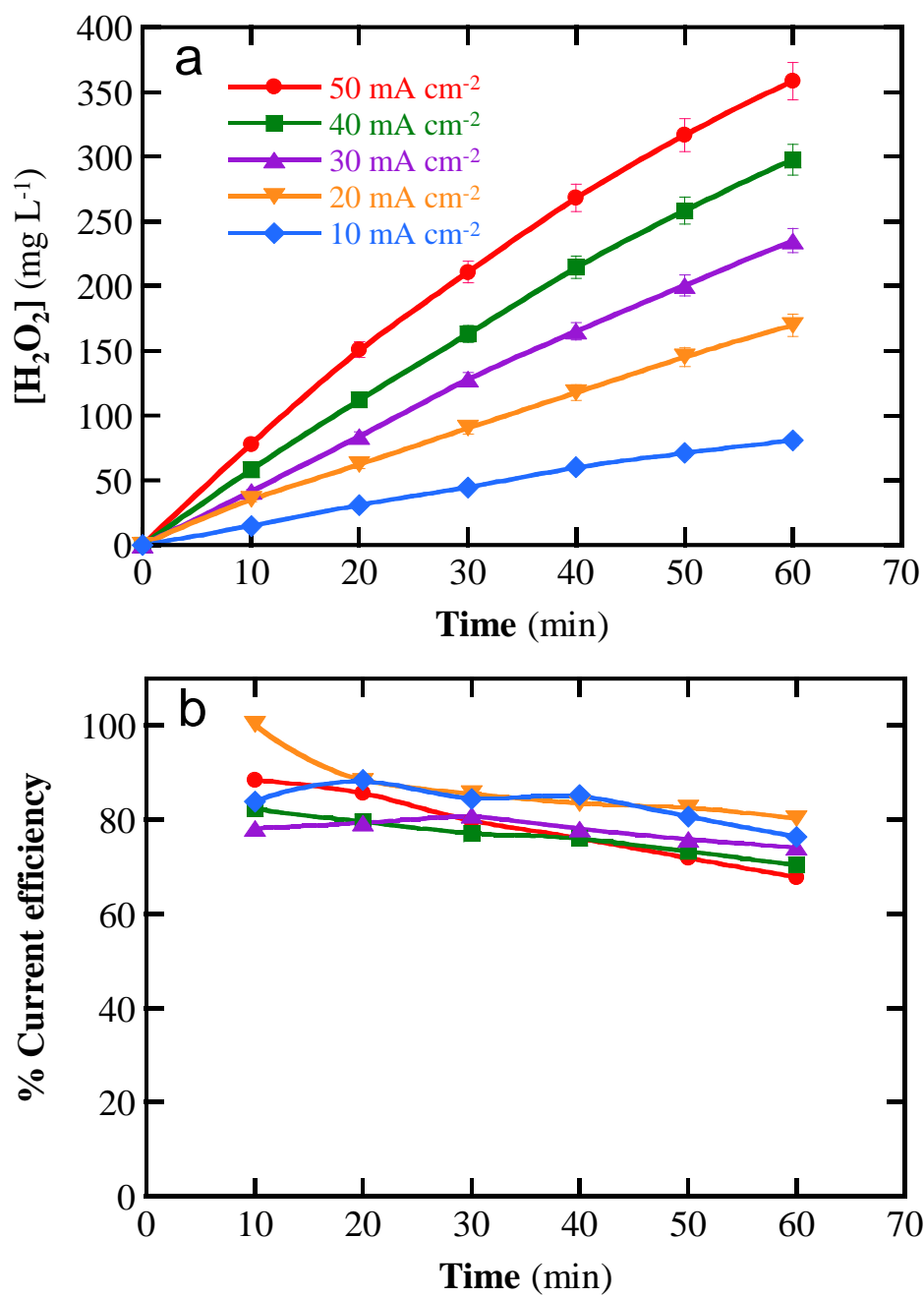

**Fig. 2**. Effect of current density on the time course of (a) the concentration of accumulated $H_2O_2$ and (b) the percent of current efficiency for 150 mL of a 0.020 M $Na_2SO_4$ at pH = 3.0 and 25 ºC using a stirred undivided tank reactor with a BDD anode and a carbon cloth air-diffusion cathode.

$$BDD + H_2O_2 \rightarrow BDD(HO_2^\bullet) + H^+ + e^- \tag{10}$$

$$2H_2O_2 \rightarrow H_2O + O_2 \tag{11}$$

The above results confirm that a carbon cloth air-diffusion electrode significantly improves the in situ electrogeneration of $H_2O_2$ by ensuring a constant supply of dissolved oxygen ($O_2$) directly at the cathode surface. The porous structure of the air-diffusion electrode then outperforms conventional 2D cathodes because it allowed an efficient $O_2$ diffusion from the air into the aqueous medium, where it is readily reduced to $H_2O_2$ via reaction (3). This continuous $O_2$ supply enhances $H_2O_2$ production rates, even at higher current densities, by maintaining optimum oxygen availability and minimizing mass transfer limitations, which is critical for the EF process [42,45].

*3.2. Understanding the impact of operational variables on Fipronil degradation*

*3.2.1. Effect of pH*

Solution pH is a crucial parameter that significantly influences the EF process because it affects the generation of $^\bullet OH$ from Fenton's reaction (2), which is responsible for pollutant removal. It also regulates the solubility and redox activity of iron species ($Fe^{3+}/Fe^{2+}$) and the stability of $H_2O_2$ [46].

Fig. 3a shows the normalized fipronil concentration vs. time plots obtained for 150 mL of 10 mg L$^{-1}$ pesticide in pure water with 0.020 M $Na_2SO_4$ and 0.50 mM $Fe^{2+}$ at the pH range 1.0-7.0 using an undivided BDD/carbon cloth air-diffusion cell at $j = 20$ mA cm$^{-2}$. As can be seen, the best performance was found at pH = 3.0, which corresponded to the optimum conditions of Fenton's reaction (2). At a more acidic medium pH = 1.0, the degradation process was strongly inhibited because of the protonation of $H_2O_2$ to form the inactive form $H_3O_2^+$ [47]. Moreover, the degradation rate gradually decreased at pH > 3.0 due to the progressive precipitation of iron hydroxides that substantially reduced

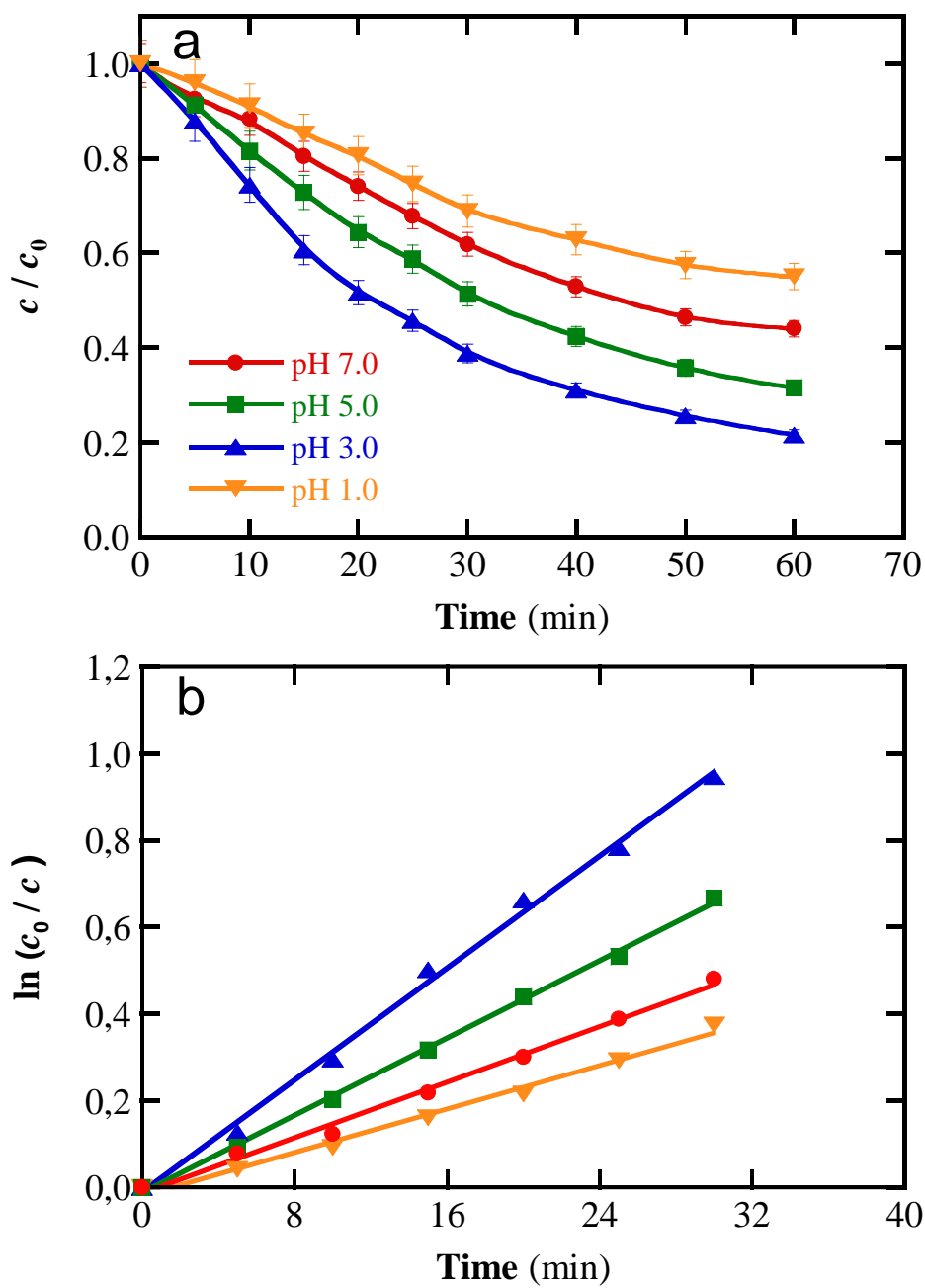

**Fig. 3**. (a) Effect of pH on the change of the normalized fipronil concentration with electrolysis time on the homogeneous EF treatment of 150 mL of 10 mg L$^{-1}$ pesticide in pure water with 0.020 M Na$_2$SO$_4$ and 0.50 mM Fe$^{2+}$ at current density ($j$) of 20 mA cm$^{-2}$. (b) Pseudo-first-order kinetic analysis of the above assays.

**Table 1.**

Pseudo-first-order rate constant, its squared linear regression coefficient, and electrical energy per order obtained for the homogeneous EF process with a BDD anode and a carbon cloth air-diffusion cathode of 150 mL of fipronil solutions in pure water with 0.020 M $Na_2SO_4$ under different experimental conditions.

| [Fipronil] (mg $L^{-1}$) | [$Fe^{2+}$] (mM) | pH | $j$ (mA $cm^{-2}$) | $k_1$ ($10^{-2}$ $min^{-1}$) | $R^2$ | $E_{EO}$ (kWh $m^{-3}$ $order^{-1}$) |
|---|---|---|---|---|---|---|
| 1.0 | 0.50 | 3.0 | 30 | 1.49 | 0.997 | 19.46 |
| 2.5 | 0.50 | 3.0 | 30 | 2.93 | 0.993 | 10.06 |
| 5.0 | 0.50 | 3.0 | 30 | 3.37 | 0.994 | 8.35 |
| 10 | 0 | 3.0 | 30 | 1.44 | 0.995 | 7.89 |
| | 0.25 | 3.0 | 30 | 3.65 | 0.994 | 7.28 |
| | 0.50 | 1.0 | 20 | 1.22 | 0.993 | 5.85 |
| | 0.50 | 3.0 | 10 | 2.86 | 0.983 | 2.13 |
| | 0.50 | 3.0 | 20 | 3.23 | 0.990 | 4.93 |
| | 0.50 | 3.0 | 30 | 4.09 | 0.998 | 6.72 |
| | 0.50 | 3.0 | 40 | 4.51 | 0.995 | 10.93 |
| | 0.50 | 3.0 | 50 | 5.62 | 0.996 | 12.80 |
| | 0.50 | 5.0 | 20 | 2.19 | 0.985 | 5.14 |
| | 0.50 | 7.0 | 20 | 1.62 | 0.996 | 5.39 |
| | 1.0 | 3.0 | 30 | 5.22 | 0.976 | 6.21 |
| 20 | 0.50 | 3.0 | 30 | 5.21 | 0.972 | 4.26 |

the concentration of $Fe^{2+}$ to catalyze the reaction (2) [48]. Despite this, it is noticeable that only a 78% abatement was reached after 60 min of electrolysis at pH = 3.0, indicating the high recalcitrance of the pesticide that slowly reacted with BDD(•OH) and homogeneous •OH.

Fig. 3b presents the excellent linear correlations found for the pseudo-first-order kinetic analysis of the data of Fig. 3a according to Eq. (7). The slope of these profiles corresponds to the pseudo-first-order rate constant $k_1$ for fipronil degradation and its value for each pH along with the squared linear regression coefficient is collected in Table 1.

The greater $k_1$ = 0.0323 min$^{-1}$ was obtained for the optimum pH = 3.0, confirming that this pH yielded the greater production of •OH from Fenton's reaction (2). The electrical energy per order calculated for the same assays from Eq. (9) is also given in Table 1. This value slightly varied with pH and progressively raised from 4.93 kWh m$^{-3}$ order$^{-1}$ at pH = 3.0 to 5.85 kWh m$^{-3}$ order$^{-1}$ at pH = 1.0.

The above findings allow the conclusion that the EF process for fipronil degradation is optimal when operating at pH = 3.0, which was chosen for subsequent trials to evaluate the effect of other operating variables. While this acidic pH is necessary to maximize the efficiency of the Fenton reaction by preventing iron precipitation and ensuring Fe$^{2+}$ availability, a pH adjustment may be required for real-world applications in natural water bodies at neutral pH, making the process adaptable to practical scenarios.

*3.2.2. Effect of current density*

The current density defines the rate of electrons delivered per unit of time and hence, largely affects the rates of H$_2$O oxidation from reaction (1) and the H$_2$O$_2$ electrogeneration from reaction (3). It is then essential to evaluate the impact of different *j*-values to identify the maximum degradation performance of the system while ensuring good energy efficiency is maintained. Fig. 4a depicts the rise of the normalized fipronil content with time for the treatment of 150 mL of 10 mg L$^{-1}$ pesticide in pure water with 0.020 M Na$_2$SO$_4$ and 0.50 mM Fe$^{2+}$ at pH = 3.0 when *j* increased from 10 to 50 mA cm$^{-2}$. While the pesticide was abated by 70% at *j* = 10 mA cm$^{-2}$ in 60 min, its degradation reached 96% at *j* = 50 mA cm$^{-2}$. The improved degradation performance at a higher *j*-value can be primarily ascribed to the greater electrogeneration of BDD(•OH) from reaction (1) and H$_2$O$_2$ from reaction (3), giving more •OH from Fenton's reaction (2), which are the key ROS for fipronil removal in the EF process [49].

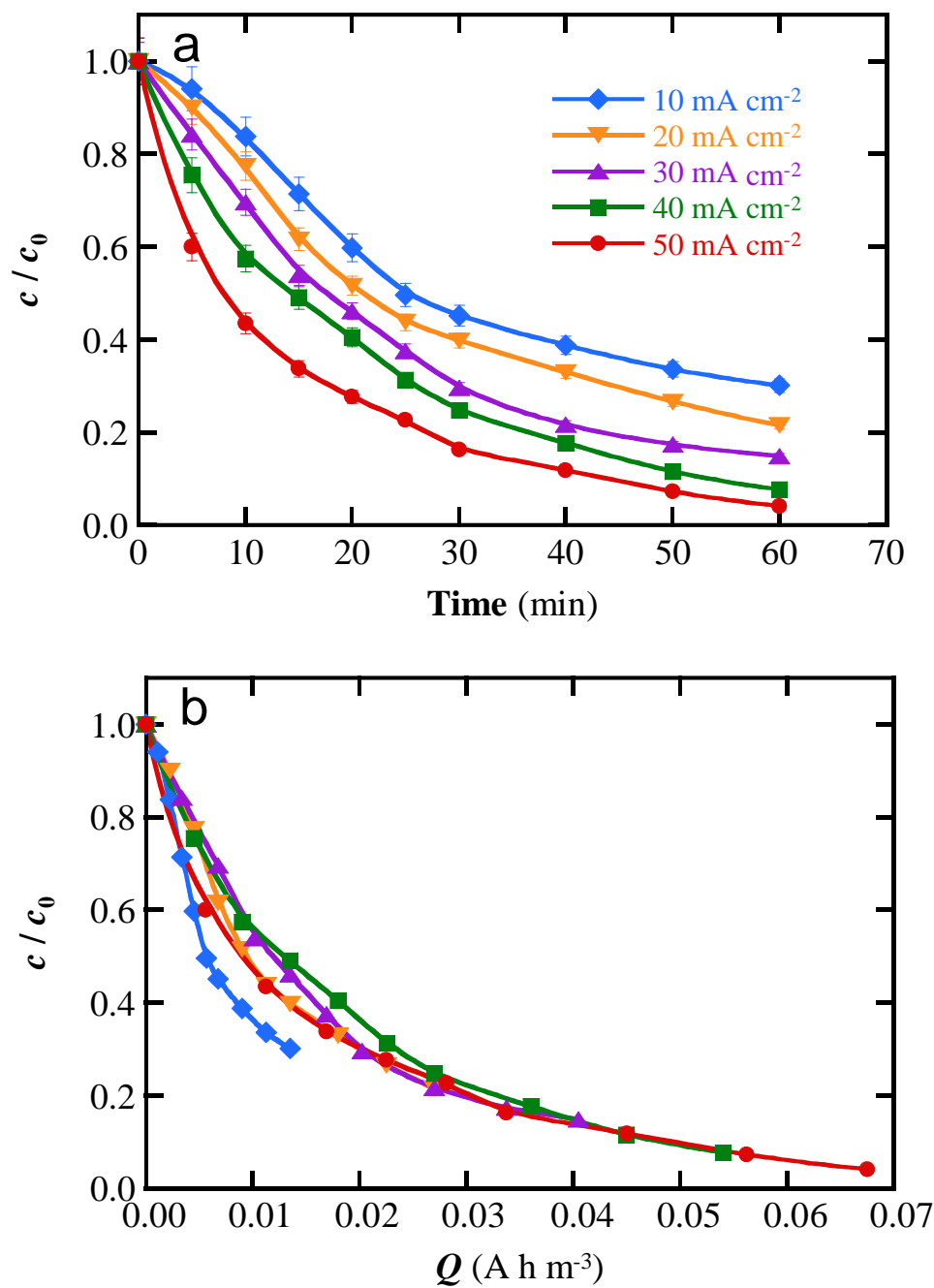

**Fig. 4**. Effect of $j$ on the variation of the normalized fipronil content with (a) electrolysis time pH and (b) the specific consumed charge for the homogeneous EF treatment of 150 mL of 10 mg L$^{-1}$ pesticide in pure water with 0.020 M Na$_2$SO$_4$ and 0.50 mM Fe$^{2+}$ at pH = 3.0 using the system of Fig. 2.

This trend was confirmed by the kinetic analysis, which also obeyed the pseudo-first-order equation (7). The corresponding $k_1$-values are listed in Table 1 and varied from 0.0286 to 0.0562 min$^{-1}$ when $j$ changed from 10 to 50 mA cm$^{-2}$. The pseudo-first-order nature of the reaction is consistent with the assumption that the concentrations of BDD(•OH) and •OH are the limiting factors of the degradation process and that both contents raised with enhancing the rate of reactions (1)-(3) [50-52]. It is interesting to remark the loss of the oxidation power of the system when raising $j$, i.e., more ROS are generated but with lower relative concentration. This suggests a stronger acceleration of the non-oxidizing parasitic reactions of the generated ROS at higher $j$. For example, these undesired reactions involve the dimerization of •OH by reaction (12) and its reaction with $H_2O_2$ to yield the weaker oxidant $HO_2^•$ by reaction (13) [44]. It is also noticeable the much greater $k_1$-values found for 10 mg $L^{-1}$ pesticide at $j = 20$ mA $cm^{-2}$ by EF as compared to the $7.2 \times 10^{-3}$ $min^{-1}$ previously determined by ECO under similar conditions [29]. In EF, the $k_1$-value was 4.5 times higher at pH = 3.0, and even 2,2 times higher at pH = 7.0, demonstrating the effectiveness of Fenton's reaction (2) to generate the oxidant •OH.

$$2•OH \rightarrow H_2O_2 \quad (12)$$

$$•OH + H_2O_2 \rightarrow HO_2^• + H_2O \quad (13)$$

Fig. 4b shows the decay of the normalized fipronil content of Fig. 4a against the specific consumed charge ($Q$). At the lowest $j = 10$ mA $cm^{-2}$, fipronil underwent a modest decay of 70% in 60 min with a moderate amount of $Q$ of 0.0135 A h $m^{-3}$, consistent with the low generation of ROS at lower currents. As $j$ was raised, a higher percentage of pesticide was removed in 60 min but with a greater $Q$ value associated with the production of more ROS content. So, 0.0270, 0.0405, 0.0540, and 0.0675 A h $m^{-3}$ were consumed for 20, 30, 40, and 50 mA $cm^{-2}$, respectively. For the two latter and greater $j$-values, the large ROS generation yielded an almost total mineralization with 93% and 96% abatement for 40 and 50 mA $cm^{-2}$, respectively. More moderate decays of 78% at 20 mA $cm^{-2}$ and 85% at 30 mA $cm^{-2}$ were found.

While the above results clearly demonstrated that higher $j$-values led to more effective fipronil degradation, a balance with the energy consumption is required to optimize the process. As pointed out above, high currents enhance undesirable parasitic reactions that do not contribute to pollutant removal, giving rise to energy loss in reactions that are not conducive to pollutant degradation and/or mineralization [53]. To clarify this point, EE/O values of 2.13, 4.92, 6.71, 10.97, and 12.80 kWh m$^{-3}$ order$^{-1}$ for 10, 20, 30, 40, and 50 mA cm$^{-2}$, respectively, were determined (see Table 1). This parameter is a crucial indicator of the energy efficiency of the EF process [54], and its progressive rise at higher $j$-values indicated a substantial increase in the energy cost, which became less favorable for the process implementation in terms of operational expenditures. The smaller EE/O values at $j$ = 10 and 20 mA cm$^{-2}$ allow inferring that good energy efficiency is achievable but at the expense of slower degradation rates (i.e., higher retention times in the reactors). Note that longer treatment times might be undesirable for treating large volumes of water where rapid treatments are desired. On the other hand, operating at $j$ = 40 mA cm$^{-2}$ only marginally improved the degradation efficiency compared to $j$ = 30 mA cm$^{-2}$, but resulted in much higher energy costs. For $j$ = 50 mA cm$^{-2}$ with the fastest degradation, the highest EE/O = 12.80 kWh m$^{-3}$ order$^{-1}$ was determined, which implied a significant energy inefficiency due to the excess of current leading to energy losses by parasitic reactions. In this context, $j$ = 30 mA cm$^{-2}$ with EE/O = 6.71 kWh m$^{-3}$ order$^{-1}$ represents a reasonable trade-off between energy efficiency and effective pollutant removal, avoiding the dramatic rise in energy consumption at higher $j$-values. From these considerations, $j$ = 30 mA cm$^{-2}$ was optimal for the subsequent trials.

*3.2.3. Effect of pesticide concentration*

The pollutant concentration is linked to the quantity of organic load that can be oxidized, which consequently affects the efficiency of the EF process. High content can help to elucidate the system robustness, with the lower values approaching environmental applications closer to real-world scenarios where trace levels of organic pollutants in contaminated water sources are usually found. The shift toward lower pollutant concentrations is crucial for advancing EAOP technologies [55].

Following the above considerations, the EF process was investigated for fipronil concentrations between 1 and 20 mg L$^{-1}$ at the optimum conditions of pH = 3.0 and $j$ = 30 mA cm$^{-2}$. The lower content of 1 mg L$^{-1}$ mimics real-world scenarios, whereas the higher concentration of 20 mg L$^{-1}$ helps to probe the limitation of the system. Fig. 5a illustrates the normalized pesticide concentration vs. time for these trials. A faster fipronil decay can be observed with increasing its content. While only 46% of the pesticide was removed in 60 min for 1 mg L$^{-1}$, a large abatement of 93% was reached for 20 mg L$^{-1}$. This tendency can be ascribed to enhanced mass transfer effects at higher organic load, which caused a higher frequency of interaction between the pollutant molecules and the generated ROS, thereby increasing the overall reaction kinetics and enhancing the degradation rate. At elevated fipronil concentrations, the increased availability of pollutant molecules leads to a higher frequency of interactions with the generated ROS, thereby accelerating the overall reaction kinetics and improving the degradation rate. Additionally, at higher concentrations, a larger proportion of ROS is utilized for direct fipronil degradation rather than being consumed in parasitic or competitive reactions with background species. In contrast, at the lowest concentration of 1 mg L$^{-1}$, the degradation efficiency is reduced due to several factors. First, the lower number of fipronil molecules decreases the probability of effective collisions and interactions with ROS, resulting in slower reaction kinetics. Second, competitive reactions, such as hydroxyl radical scavenging by background organic matter or other intermediates, become more significant at lower pollutant concentrations, further reducing the availability of ROS for fipronil degradation.. Note that the EF process is still effective for the lowest concentration of 1 mg L$^{-1}$ but with much smaller pollutant removal as compared to higher concentrations because of its smaller number of collisions and interactions with the ROS. For 10 and especially 20 mg L$^{-1}$, the pollutant removal remained relatively efficient due to a more optimum balance between the mass transfer and reaction kinetics. This behavior is consistent with observations in other advanced oxidation processes, where lower pollutant concentrations often result in reduced degradation rates due to kinetic and competitive effects [29,56, 57].

A pseudo-first-order reaction was followed by all the assays of Fig. 5a. The corresponding $k_1$-values given in Table 1 progressively increased from 0.0149 min$^{-1}$ at 1 mg L$^{-1}$ to 0.0521 min$^{-1}$ at 20 mg L$^{-1}$ according to the enhancement of the process at higher fipronil content. The same behavior was found for the corresponding EE/O values that showed a clear inverse relationship with the pollutant concentration related to more efficient utilization of generated ROS. Then, 19.46, 10.06, 8.35, 6.72, and 4.25 kWh m$^{-3}$ order$^{-1}$ were determined for 1, 2.5, 5, 10, and 20 mg L$^{-1}$ fipronil. This tendency highlights that the mass transfer limitations at lower concentrations not only significantly reduce the degradation efficiency but also significantly increase the energy consumption.

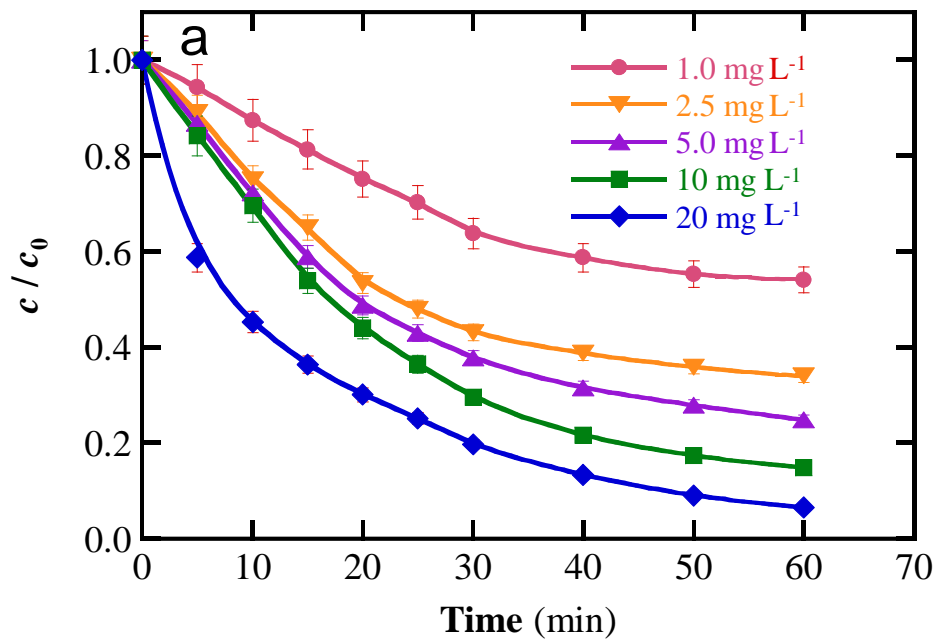

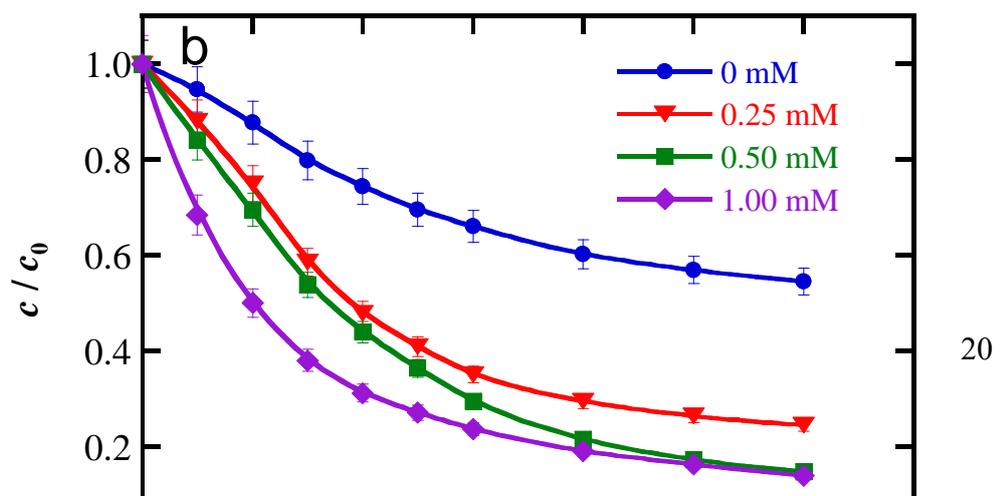

**Fig. 5**. Effect of (a) fipronil concentration at 0.50 mM Fe$^{2+}$ and (b) Fe$^{2+}$ dosage at $j$ = 30 mA cm$^{-2}$ over the normalized fipronil content vs. electrolysis time pH for the homogeneous EF process of 150 mL of 10 mg L$^{-1}$ pesticide with 0.020 M Na$_2$SO$_4$ at pH = 3.0 using the system of Fig. 2.

The above findings indicate that the EF process efficiently performs through a wide range of fipronil concentrations, where the higher organic loads improved overall degradation rates. Notably, the process remained highly effective even at very low concentrations of fipronil from 1 to 5 mg L$^{-1}$, which are more representative of environmental conditions found in natural water bodies. This points the potential of the EF process for real-world applications, where pesticide concentrations are typically much lower than those used in laboratory-scale optimization studies. The ability to achieve significant degradation at environmentally relevant concentrations underscores the practical applicability of this treatment method.

### 3.2.4. Effect of Fe$^{2+}$ dosage

The main characteristic of the EF process is the addition of Fe$^{2+}$, which plays a pivotal role in generating homogeneous •OH from Fenton's reaction (2). Insufficient Fe$^{2+}$ can limit the generation of •OH, but its excess yields a waste effect from reaction (14), decreasing the process efficiency [58].

$$Fe^{2+} + {}^\bullet OH \rightarrow Fe^{3+} + OH^- \quad (14)$$

Fig. 5b depicts the effect of the concentration of the catalytic $Fe^{2+}$ over the time course of the normalized fipronil decay for 150 mL of 10 mg L$^{-1}$ pesticide with 0.020 M $Na_2SO_4$ at pH = 3.0 and $j$ = 30 mA cm$^{-2}$. In the absence of $Fe^{2+}$ (0 mM), the ECO process took place with only a 45% pesticide decay in 60 min. The lower oxidative capability of the ECO process is due to the only action of the oxidizing agent BDD(•OH) formed from reaction (1) [59]. When $Fe^{2+}$ was added, the EF process yielded a quicker degradation because of the additional •OH generation from Fenton's reaction (2). For 0.25 mM $Fe^{2+}$, the pesticide removal increased to 75% in 60 min, which reached 85% for 0.50 mM $Fe^{2+}$ due to the acceleration of such reaction with the production of more •OH. The same phenomenon can be seen in Fig. 5b for 1.00 mM $Fe^{2+}$ at the beginning of the assay, but the degradation was gradually decelerated giving rise to a similar decay of 86% in 60 min. All the concentration decays also obeyed a pseudo-first-order kinetics yielding increasing $k_1$-values of 0.0144, 0.0365, 0.0409, and 0.0522 min$^{-1}$ for 0, 0.25, 0.50, and 1.00 mM $Fe^{2+}$, respectively (see Table 1). The opposite trend was determined for their EE/O with decreasing values of 7.89, 7.28, 6.72, and 6.21.

The above findings suggest that the excess of •OH produced by 1.00 mM with respect to 0.50 mM $Fe^{2+}$ was preferentially consumed by the parasitic reaction (12) to yield a similar degradation in 60 min. Although 1.00 mM $Fe^{2+}$ yielded a slightly greater $k_1$ and lower EE/O, it did not significantly improve the performance with respect to 0.50 mM $Fe^{2+}$, which can be taken as optimal for the homogeneous EF treatment of fipronil.

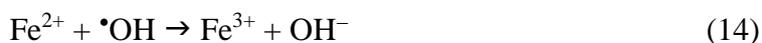
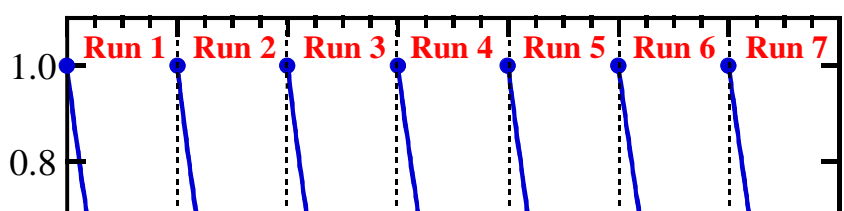

**Fig. 6**. Reusability of the system in seven consecutive degradation runs of homogeneous EF of 150 mL of 10 mg L$^{-1}$ fipronil solutions with 0.020 M Na$_2$SO$_4$ and 0.50 mM Fe$^{2+}$ at pH = 3.0 using the system of Fig. 2 at $j$ = 30 mA cm$^{-2}$.

*3.2.5. System sustained operation using reproducibility under continuous use to prove resilience*

The reproducibility of the EF process is important to ensure its performance at long electrolysis time for feasible scaled-up at the industrial level. Seven consecutive runs of 60 min were made for 150 mL of 10 mg L$^{-1}$ pesticide with 0.020 M Na$_2$SO$_4$ under the optimum conditions of 0.50 mM Fe$^{2+}$, pH = 3.0, and $j$ = 30 mA cm$^{-2}$. Fig. 6 shows a similar decay of the normalized fipronil content in 84-86% at the end of all runs. Hence, one can infer the excellent reproducibility and stability of the system in successive cycles, highlighting the robustness of the homogeneous EF process [60]. This stability suggests that the BDD anode and the carbon cloth air-diffusion cathode exhibited good durability and that the electrochemical environment (i.e., H$_2$O$_2$ production, BDD($^\bullet$OH) and $^\bullet$OH generation, and Fe$^{2+}$ cycling) remained effective over time. The lack of significant variation in degradation rate also implies that issues such as electrode fouling or iron catalyst deactivation are minimal, thus supporting the suitability of this approach for long-term pollutant degradation in real-world applications.

*3.2.6. Identification and time course of degradation by-products*

It is well-known that the final degradation of aromatic pollutants by EAOPs leads to a mixture of short-linear aliphatic carboxylic acids. To confirm this behavior for fipronil under EF treatment, the 10 mg $L^{-1}$ pesticide solutions electrolyzed under optimum conditions were analyzed by ion-exclusion HPLC. Fig. 7 presents the evolution of the 5 final carboxylic acids detected. Acetic and fumaric acids are expected to be formed from the cleavage of the aromatic moieties of the intermediates. They are oxidized to oxalic and formic acids as ultimate by-products since both compounds are directly mineralized to $CO_2$ [61,62]. In parallel, oxamic acid is expected to be produced from the degradation of N-intermediates. All these acids form Fe(III) complexes that are hardly oxidized by BDD($^\bullet$OH) and homogeneous $^\bullet$OH. As can be seen, the concentration of acetic acid remained practically constant during all the treatment, near 2.5 mg L$^{-1}$, because its generation and Fe(III)-acetate removal were balanced. Fumaric acid increased throughout all the treatments up to 8.09 mg L$^{-1}$, indicating that its production was much faster than the degradation of Fe(III)-fumarate complexes [61]. Initially, formic acid raised sharply within the first 10 min, reaching a maximum of 11.00 mg L$^{-1}$, which was further reduced to nearly 2.10 mg L$^{-1}$ between 40 and 60 min because of the mineralization of the corresponding Fe(III)-formate complexes [62]. In contrast, the content of oxalic acid showed a continuous increase attaining 39.94 mg L$^{-1}$ at 60 min. This suggests a large stability of the generated Fe(III)-oxalate species in front of the attack of hydroxyl radical. The persistence and great accumulation of oxalic acid then explain that the overall mineralization of fipronil by EF is not feasible.

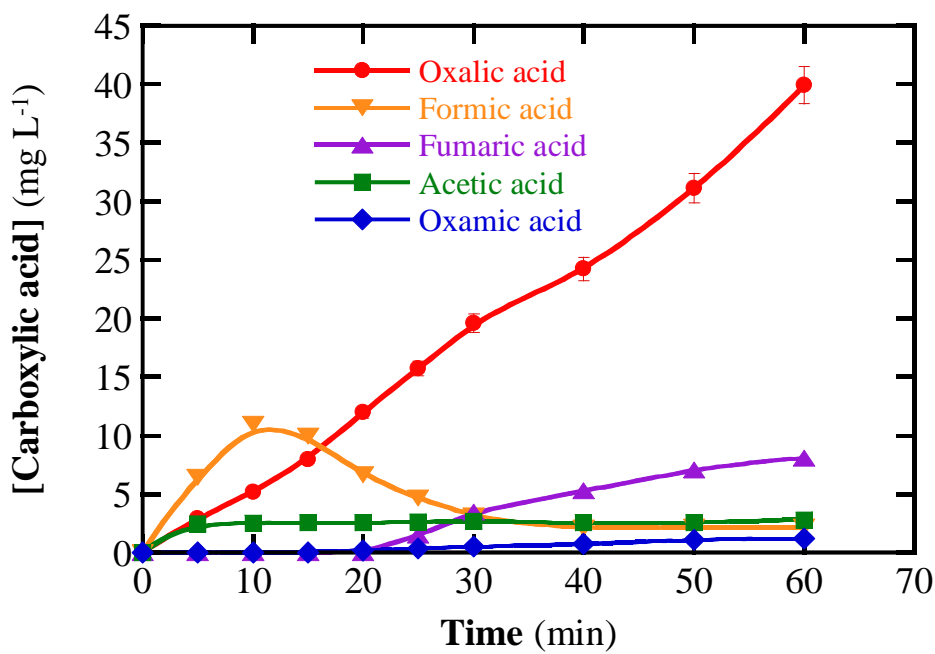

**Fig. 7.** Evolution of the concentration of the final carboxylic acids detected during the first run of Fig. 6.

Finally, oxamic acid appeared at 25 min and only reached 1.23 mg L$^{-1}$ at the end of the experiment, indicating that it was slowly accumulated in the process by the persistence of its Fe(III) complexes [63].

*3.2.7. Time-course of inorganic ions released during electrochemical treatment*

The degradation of fipronil through the EF process is accompanied by the release of inorganic ions, namely, fluoride, chloride, and ammonium. Fig. 8 highlights the time course of the concentration of such ions for the first run of Fig. 6. A continuous and significant increase in F$^-$ concentration with time can be observed, from 0.1 mg L$^{-1}$ at 5 min to 2.25 mg L$^{-1}$ after 60 min, corresponding to an 86.3% of its initial content in solution. This trend indicates the progressive cleavage of the C-F bonds in the fipronil molecule by hydroxyl radicals, although no total F$^-$ removal was achieved within the time tested. In contrast, the concentration of Cl$^-$ ion exhibits a fluctuating pattern, varying between 0.73 mg L$^{-1}$ at 5 min and 1.05 mg L$^{-1}$ at 30 min, whereupon slightly decreased to 0.85 mg L$^{-1}$ at 60 min related to 52.5% of its initial content. This behavior suggests the continuous release of Cl$^-$ ion from the C-Cl bonds of the pesticide and its partial oxidation to the strong oxidant active chlorine at the anode that can assist in organics degradation [62]. The NH$_4^+$ concentration slightly raised throughout the reaction, only reaching 0.146 mg L$^{-1}$ at 60 min. associated with 11.4% of its initial value. This suggests that only a small proportion of nitrogen-containing intermediates are being transformed into ammonium. This behavior can be attributed to the degradation pathways of aromatic primary amines, where the non-selective ·OH preferentially attacks the more nucleophilic position, influenced by the electronegative nature of the nitrogen substituent, ultimately forming ammonium ions. These findings underscore the intricate mechanisms of ions released and the effectiveness of the EF process in breaking down the molecular structure of fipronil. Since no other nitrogen-containing species like NO$_3^-$ ion was detected, it seems feasible the formation of N$_x$O$_y$ volatile compounds as found for other N-aromatics [62].

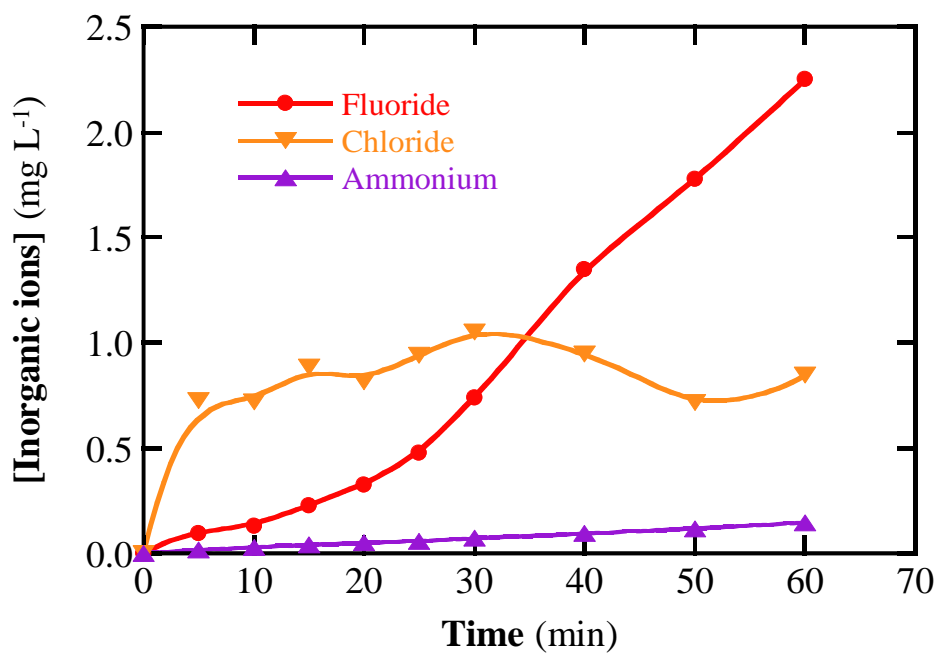

**Fig 8**. Evolution of the concentration of released inorganic ions detected during the first run of Fig. 6.

*3.2.8. Effect of scavenging*

Fig. 9 demonstrates the scavenging effect of *tert*-butanol and *p*-benzoquinone on the electro-Fenton degradation of fipronil, highlighting the role of reactive oxygen species in the process. The results reveal that the addition of tert-butanol, a hydroxyl radical ($^{\bullet}$OH) scavenger, significantly hindered the reaction, reducing the fipronil removal efficiency from 85% to 19% after 60 minutes. In contrast, the presence of p-benzoquinone, a scavenger of hydroperoxyl ($HO_2^{\bullet}$) and superoxide ($O_2^{-\bullet}$) radicals, only slightly decreased the removal efficiency from 85% to 80% over the same period. These findings strongly suggest that $^{\bullet}$OH are the primary reactive species responsible for fipronil degradation in the electro-Fenton system, while the contribution of $HO_2^{\bullet}/O_2^{-\bullet}$ radicals is relatively minor [64].

**Fig 9**. Scavenging effect of *tert*-butanol and *p*-benzoquinone in the homogeneous EF of 150 mL of 10 mg L$^{-1}$ fipronil solutions with 0.020 M Na$_2$SO$_4$ and 0.50 mM Fe$^{2+}$ at pH = 3.0 and $j$ = 30 mA cm$^{-2}$.

## 4. Conclusions

Optimization of the degradation of fipronil solutions via the EF process using a BDD as the anode and a carbon cloth air-diffusion electrode as the cathode has been made focusing on key operating parameters such as pH, $j$, initial pollutant concentration, and Fe$^{2+}$ dosage. The cathode showed a great selectivity to originate H$_2$O$_2$ from O$_2$ reduction, with current efficiencies between 80 and 90% for $j$ = 10-50 mA cm$^{-2}$. In all cases, the pesticide concentration obeyed a pseudo-first-order kinetics. The optimum pH was determined to be 3.0, consistent with the process control by Fenton's reaction (2), producing homogeneous $^{\bullet}$OH as the main oxidant. A $j$ = 30 mA cm$^{-2}$ was established as the best current density because it gave a reasonable trade-off between energy efficiency and effective pollutant removal. Fipronil degradation was enhanced by raising its concentration up to 20 mg L$^{-1}$, even being effective for environmentally relevant contents as low as 1 mg L$^{-1}$. A concentration of 10 mg L$^{-1}$ was taken as more suitable to characterize the EF treatment. The rise of Fe$^{2+}$ dosage indicated a clear enhancement of fipronil abatement due to the greater $^{\bullet}$OH production via Fenton's reaction (2) with respect to BDD($^{\bullet}$OH) generation at the anode. An optimum content of 0.50 mM Fe$^{2+}$ was found, ensuring an efficient $^{\bullet}$OH generation without significant wastage due to excessive Fe$^{2+}$. The process showed excellent reproducibility and stability across 7 consecutive optimized runs in which 85% removal was found for 10 mg L$^{-1}$ fipronil with 0.50 mM Fe$^{2+}$, pH = 3.0, and $j$ =30 mA cm$^{-2}$. This highlights the robustness and reliability of the optimized EF system, making it a viable option for sustainable pollutant removal in water treatment applications. Finally, the evolution of five carboxylic acids, including acetic, fumaric, formic, oxalic, and oxamic acids, as final by-products, along with that of released ions like F$^{-}$, Cl$^{-}$, and NH$_4^{+}$, was determined.

**CRediT authorship contribution statement**

**Caio Machado Fernandes:** Conceptualization, Methodology, Formal analysis, Investigation, Validation, Data curation, Visualization, Funding acquisition, Writing – original draft, Writing – review & editing. **Gabriel A. Cerrón-Calle:** Validation, Formal analysis, Investigation, Writing – review & editing. **Enric Brillas:** Formal analysis, Writing - Original Draft, Visualization, Writing –review & editing. **Mauro C. Santos:** Conceptualization, Supervision, Funding acquisition, Writing – review & editing. **Sergi Garcia-Segura:** Conceptualization, Methodology, Resources, Writing - Original Draft,

Supervision, Project administration, Funding acquisition, Writing – review & editing, Supervision.

**Acknowledgments**

The authors would like to thank Fundação de Amparo à Pesquisa do Estado de São Paulo (FAPESP, #2022/10484-4, #2024/03549-8, #2022/15252-4, #2022/12895-1) and Conselho Nacional de Desenvolvimento Científico e Tecnológico (CNPq, #308663/2023–3, #402609/2023–9) for the financial support.